\documentstyle[12pt,leqno]{article}
 \newcommand{\beq}{\begin{equation}}                       
 \newcommand{\eeq}{\end{equation}}                         
 \newcounter{nt}[section]                                  
 \newcounter{nl}[section]                                  
 \date{}                                                   

\begin{document}

 \title{ \bf  On the FitzHugh - Nagumo Model }

\vspace{15mm}

\author{\sc  M. De Angelis, P.Renno
   \thanks{  Faculty of Engineering, Department of Mathematics and Applications, Naples, Italy. e-mail modeange@unina.it }}

 \maketitle

 \begin{abstract}
The initial value problem ${\cal P}_0,\,$ in all of the space, for the spatio - temporal FitzHugh - Nagumo equations is analyzed. When the reaction kinetics of the model can be outlined by means of piecewise linear approximations, then the solution of ${\cal P}_0\,$ is explicitly obtained. For periodic initial data are possible damped travelling waves and their speed of propagation is evaluated. The results imply applications also to the non linear case.
\end{abstract}

Keywords:    {\em Parabolic systems. Biological models. Laplace transforms}
  \vspace{7mm}

 \section{Introduction}
 \setcounter{equation}{0}

\vspace{7mm}

\hspace{5.7mm}One of the reaction  diffusion systems which models various important biological phenomena is given by 

	\vspace{3mm}
  \beq                                                     \label{11}
  \left \{
   \begin{array}{lll}
    \frac{\partial \,u }{\partial \,t } =\,  \varepsilon \,\frac{\partial^2 \,u }{\partial \,x^2 } \,-\, v\,\,  + f(u ) \,  \\
\\
 \frac{\partial \,v }{\partial \,t } \, = \, b\, u\, - \beta\, v\, 
\\
   
   \end{array}
  \right.
 \eeq

\vspace{3mm} \noindent where the appropriate class of functions $\, f(u)\,$ depends on the reaction kinetics of the model \cite{i} - \cite{ks}. In the theory of nerve  membranes, for example,  the system (\ref{11}) is related to the spatio-temporal FitzHugh - Nagumo equations (FHN) with

\vspace {3mm}
\beq                 \label{12}
f(u)= u\, (\, a-u \,) \, (\,u-1\,) \,\qquad\qquad (\,0\,<\,a\,< \,1\,)
\eeq

\vspace {3mm}\noindent
and where  $\, u\,( \,x,t\,)$  models the transmembrane voltage  of a nerve axon at distance x and  time t, $\, v\,(\,x,t\,)$ is an auxiliary variable that  acts as recovery variable. Further, the  diffusion coefficient $\, \varepsilon \,$  and the parameters  $  \,  b, \, \beta \,$ are  all non negative  \cite {kss},\cite {f61}.

In addition to the propagation of nerve action potentials, ({\ref{11}) - (\ref{12}) govern other several  biological and biochemical  phenomena; the list of references is very long and the variety of analytical aspects examined is wide. \cite {wzd} - \cite{nya}. Further,we must remark travelling pulses and periodic wavetrains obtained by means  of piecewise linear approximations of $\, f(u)\,$ as

\vspace {3mm}
\beq                 \label{13}
f(u)= \,\,\eta \, (\, u-a \,) \, \,-u\, \,\,\,\qquad\qquad\, \ (\,0\,<\,a\,< \,1\,)
\eeq

\vspace {3mm}\noindent where $ \,\eta \,$ denotes the unit- step function.\cite {mk}- \cite{t}.

\vspace{3mm}
Typical boundary value problems related to the linear case (\ref{13}) can be explicitly solved. Aim of this paper is the analysis of the initial value problem ${\cal P}_0\,$  in all of the space; the fundamental solution and the explicit solution of ${\cal P}_0\,$  are determined when (\ref{13}) holds. More, in the non linear case of the FHN model given by (\ref{11})-(\ref{12}), the problem ${\cal P}_0\,$  is reduced to appropriate integral equations whose kernels are functions characterized by basic properties. All this implies existence and uniqueness properties, together with a priori estimates.

  \vspace{5mm}

 \section{Statement of the problem and results }
 \setcounter{equation}{0}

\vspace{7mm}

 \hspace{5.7mm}  Both the non linear source  (\ref{12}) and the linear approximation (\ref{13}) involve a linear term $\,\, -\, k\,u\,\,\,$  with $ k\,=\,a\,$ for (\ref{12}), and $ \, k\,=\, 1\, $  for (\ref{13}). As consequence, the system (\ref{11}) becomes

\vspace{3mm}
  \beq                                                     \label{21}
  \left \{
   \begin{array}{lll}
    {u_t } \,-  \varepsilon \, u_{xx} + k\,u + \,v \,   = \, \varphi (u)  
\\    \hspace{7cm}  (x,t)\in \Re\\
 v_t  \, + \beta \, v \, - b\, u\,=\,0 \,\,\, \,,
\\
   
   \end{array}
  \right.
 \eeq 

\vspace{3mm}\noindent where $\,\varphi \,(u)\, = \,\,u^2\, (\, a\,+\,1\,-\,u\,)\,\,$ when $ \,f\,$  is given by (\ref{12}), while, in the linear case, 
$\,\varphi \,(u)\,$ is equal to the constant $\,\,\bar \eta \,\, $ that holds zero or one.

\vspace{3mm}The initial- value problem  $\,{\cal P}_0$  related to (\ref{21}) with $\,\varphi \,(u)\,= \,\bar \eta \,\, $ is analyzed in the set 

\vspace{3mm}
\beq    \Omega_T =\{(x,t) :  x \in \Re
, \  \ 0 < t \leq T \},
\eeq

\vspace{3mm}\noindent with the conditions 

\vspace{3mm} 
   \beq                                     \label{23}                                                
u(x,0)= u_0(x), \  \  \ \  v(x,0)=v_0(x), \  \ \ \\\ \ \,\,\\ \,\,\, x \in \Re.  
\eeq

\vspace{3mm}When the linear approximation of $\, f\,$ holds, then the explicit solution of the problem ${\cal P}_0\,$ can be obtained  by means of functional transforms (Fourier with respect to $\, x\,$ and  Laplace as for $\, t \,$). If one puts formally 

\vspace {3mm}
\beq     \label{24}
\hat u (x,s) \, = \int_ 0^\infty \,\, e^{-st} \,\, u(x,t) \,dt \,\,, \,\,\,  \,\hat v (x,s)   \, = \int_ 0^\infty \,\, e^{-st} \,\, v(x,t) \,dt \,,\,
\eeq

\vspace {3mm}\noindent from (\ref{21}) (\ref{23}) one deduces:

\vspace{3mm}

\beq    \label{25}
{\hat u }=  \int_\Re  [\, u_0(\,\xi\,)\, +\,\frac{\bar\eta}{s}\,\,]\,\,  \hat K_0 \,[(x\,- \xi, s\,)\, \,d\xi - \int_\Re v_0 (\xi)\,\hat K_1 \,(\, x-\xi, s\,)\,\,d\xi
\eeq

\beq    \label{26}
{\hat v } = b \int_\Re   [ u_0(\xi) + \frac{\bar\eta}{s} ]\hat K_1 ( x-\xi, s) d\xi -  b \int_\Re   v_0  (\xi) \hat K_2( x-\xi, s) d\xi+ \frac{ v_0  (x)}{s+\beta},
\eeq

\vspace {3mm}\noindent where 

\vspace{3mm}
\beq                 \label{27}
\hat K_n (x,s) = \,\, \frac{e^{- \frac{|x|}{\sqrt \varepsilon} \,\,\sigma}}{2 \,\, \sqrt\varepsilon \, \,\,\sigma \, (s + \beta )^n}  \ \ \ \,\,\,\,\,\,\, (\,n=0,1,2\,)
\eeq

\vspace{3mm} \noindent with  $ \,\,\,\, \sigma^2 \ \,=\, s\, +\, k \, + \, \frac{b}{s+\beta}.\,\,$

 \vspace{3mm}

Now, let us consider the fundamental solution

\vspace{3mm}
\beq     \label {28}
g( x,y) \, = \, \frac{1}{2 \, \sqrt{\pi \varepsilon y}} \,\,e^{ \,-\,k\,y\,-\, \frac{x^2}{4 \varepsilon  y}}
\eeq

\vspace{3mm}\noindent of the heat equation $\, \varepsilon \, g_{xx}\, -\, g_y\, - k\,g\,=\,0\,\,$ and,  for  $\, n=0,1,2,\,\,\, $let

\vspace{3mm}
\beq     \label {29}
G_n( x,t)  =  \int^t_0  e^{-\beta (t-y)}  g(x,y) \Bigl( \frac{t-y}{by}\Bigl)^{\frac{n-1}{2}}\,J_{n-1} \,\Bigl(2 \sqrt{b(t-y)y} \,\Bigl )dy,
\eeq

\vspace{3mm} \noindent where  $ J_n (z) \,$    denotes the Bessel function of first kind. Then, if one puts

\vspace{3mm}
 \beq    \label{210}
 K_0 \, =\, \, g(x,t) + G_0(x,t) ,\,\,\,\,\,\,\, K_i \,=\, G_i(x,t) \ \,(i=1,2),   
\eeq

\vspace{3mm}\noindent the following theorems hold.

\vspace{5mm}

{\bf Theorem 2.1}- {\em In the half-plane} $ \Re e  \,s > \,max(\,-\,k ,\,-\beta\,),\,$   {\em the Laplace integrals} of  $\,\,K_n (x,t)\, \,(n=0,1,2)\,$ {\em  converge absolutely for all  }$\,|x|>0,\,$  {\em and one has}$\,\,\,\,
{\cal L}_t \, K_n (x,t)\ =\, \hat K_n (x,s).\,$ 

\vspace{5mm}
{\bf Theorem 2.2}- {\em The functions}  $\  K_0, \, K_1,\, K_2 \, $ {\em are $\, C ^ {\,\infty}\,( \Omega_T)\,\,$ solutions of  the integro differential equation  }:

\beq     \label{211}
 {z_t } \,-  \varepsilon \, z_{xx} + k\,z  + b \,\int^t_0 \, e^{- \beta (t-\tau)} \,\,z(x,\tau) \, d\tau \,\, =\,\,0  
\eeq

\vspace{3mm}\noindent  {\em  and have the same basic properties of the fundamental solution } (\ref{28}) {\em of the heat operator. Further, for} $\, i=\,0\,${\em and} $i\,=1,\,$ {\em it results}:

\vspace{3mm}
\beq     \label {212}
K_i(x,t)\,=\, (\, \partial _t \,+\beta )\, K_{i+1}, \, \,\,\,\,\,\,\,  \lim_ {t\downarrow 0}K_{i+1}\,=\,0 \,\, \,\,(\,i=\,0,1)
\eeq

\vspace{3mm} \noindent{\em  while} $ \displaystyle {\lim_ {t\rightarrow 0}}\,\,K_0\,(x,t\,)=\,0\,\,$ {\em only for }$\,|x|\,>0$.

\vspace{3mm}These properties assure the convergence of the convolutions

\beq    \label{213}
K_n\,* \, \psi \, = \, \int _\Re \, K_n(\,x-\xi,t\,) \, \,\psi(\xi ) \ d \xi\,\,\,\, \,\,(n=0,1,2)
\eeq

\vspace{3mm}\noindent for all the functions that satisfy a {\em growth condition }of the form 

\vspace{3mm}
  \beq             \label{214}                                        
    |\psi(x)|\, < \, c_1\, exp\, [ \, c_2 \, |x|^{\alpha +1}\,], \,\, \;\, 0\,< \alpha \, < \,1 
\eeq  

\vspace{3mm} \noindent with $\, c_1\, $ and  $ \,c_2\,$  positive constants.

 \vspace{3mm} Further, let consider  the following functions  

\vspace{3mm}

\beq   \label{215}
N_1(t) = \frac{\bar \eta \beta }{b+\beta}\Bigl[1- e^{-\frac{1+\beta }{ù2}t }\cos{(\gamma  t )}\Bigl] +\frac{\bar \eta(2b+\beta-\beta^2)}{2\gamma (b+\beta)}\, e^{-\frac{1+\beta }{2}t } \sin (\gamma t)
\eeq

\vspace{3mm}

\beq   \label{216}
N_2(t) = \frac{\bar \eta  }{b+\beta}\Bigl[1- e^{-\frac{1+\beta }{2}t }\cos{(\gamma  t )}\Bigl] -\frac{\bar \eta(1+\beta)}{2\gamma (b+\beta)}\, e^{-\frac{1+\beta }{2}t } \sin (\gamma t),
\eeq

\vspace{3mm}\noindent where $\, \gamma \,= \Bigl[\, b\,-\, (\, \beta -1 \, )^2 /4\,\Bigl]^{1/2}.$ Then, by (\ref{25})-(\ref{26}) and the foregoing statements, the explicit solution of the linear problem ${\cal P}_0\,$is given by

\vspace{3mm}
\beq            \label{217}
  \left \{
   \begin{array}{lll}
    u =  \, N_1 \,(t) \, + \,u_0\, * \, K_0\, -\, v_0 \, *\, K_1
  \\
\\

  v  =  \, N_2 \,(t) \, + \,u_0\, * \, K_1\, -\, v_0 \, *\, K_2\,+\, v_0\,(x)\, e^{\,-\, \beta \, t\,}, 
   \end{array}
  \right.
 \eeq

\vspace{3mm}\noindent and the following conclusion is deduced.

\vspace{5mm} 
{\bf Theorem 2.3} - {\em When the data }$\,(\, u_0 ,\,\,v_0 )\,\,$ {\em are continuous functions that satisfy the  growth condition } (\ref{214}), {\em  then the formulae }(\ref{217})  {\em represent the unique solution of the problem ${\cal P}_0\,$ in the class of solutions compatible with  }(\ref{214}).
\vspace{5mm}

 \section{Travelling waves and a priori estimates }
 \setcounter{equation}{0}

\vspace{7mm}

 \hspace{5.7mm}

\vspace{3mm}A first example of applications is related to the linear case  and concerns the analysis of travelling waves. By the explicit solution (\ref{217}),  for instance, when $\, \bar \eta \,=\,0,\,\, v_0,\, =0\,\,\, u_0\,= A\, cos (w\,x),\,$ then one obtains:

\vspace{3mm}
\beq   \label{31}
u\,=\, (\, \partial _t\, +\, \beta\, )w,\,\,\,\,\, v\,= b\,w
\eeq

\vspace{3mm}\noindent where 

\vspace{3mm}
\beq   \label{32}
w\,=\,\frac{A}{2\,\alpha} \,e^{\,-\,\frac{1+\beta+\varepsilon w^2}{2}\,t\,}\,\, \Bigl [\,\sin (\, wx\,+\,\alpha t)\, - \,\sin ( \, wx\,-\,\alpha t)\,\Bigl ]\, 
\eeq

\vspace{3mm}\noindent with  $ \, \alpha \,= \Bigl[\, b\,-\, \Bigl( \frac{1+\varepsilon w^2 -\beta }{2}\Bigl)^2\,\Bigl]^{1/2}.\,\,$ So, when $ \, b\,>\, \Bigl( \frac{1+\varepsilon w^2 -\beta }{2}\Bigl)^2,\,\,\,$ there exist damped travelling waves with speed equal to $\, \alpha / w.\,\,$

\vspace{3mm}Moreover, in the non linear case, by (\ref{25}),(\ref{26}) one deduces the following integral equations 

\vspace{3mm}
\beq            \label{33}
  \left \{
   \begin{array}{lll}
    u =  \, \,u_0\, * \, K_0\, -\, v_0 \, *\, K_1\,\,+ \int _0^t \, \varphi (u) \,* K_0 \,\ d\tau
  \\
\\

  v  =  \,b \,u_0\, * \, K_1\, -\,b\, v_0 \, *\, K_2\,+\, v_0\,(x)\, e^{\,-\, \beta \, t\,} \,+ \,b\, \int _0^t \, \varphi (u) \,* K_1 \,\ d\tau 
   \end{array}
  \right.
 \eeq

\vspace{3mm}\noindent that imply a priori estimates. In fact, in the class of bounded solutions, if one puts

 \vspace{3mm}
\beq     \label{34}
   ||u_0|| = \sup_{ x\in \Re}  |u_0(x)| ,\,\,\,||v_0|| \,=\, \sup_{ x\in \Re} \, | v_0(x)| , \,\,\,  ||\varphi|| = \sup_{ u\in \Re}  |\varphi(u)| 
\eeq

\vspace{3mm}\noindent  by means of the basic properties of the kernels $ \, K_0,\,K_1\, K_2,\,$ one obtains:

\vspace{3mm}
\beq            \label{35}
  \left \{
   \begin{array}{lll}
   | u | <  \, \,c_1\,||\varphi||\,+\, c_2 \,(\,\,||u_0||\,t \, +||v_0||\,) \,E(t)
  \\
\\

  |v| <    \, \,c_1\,||\varphi||\,+\, c_2 \,(\,\,||u_0||\, \, +||v_0||\,t\,\,) \,E(t)
   \end{array}
  \right.
 \eeq

\vspace{3mm}\noindent where 

\vspace{3mm}
\beq   \label{36}
 E(t)\,= \frac{e^{\,-\,k\,t\, }- e^{\,-\,\beta\,t\, }}{\beta-k}
\eeq

\vspace{3mm}\noindent and where the constants $\,c_1,\,c_2\,$ depend on the parameters $\, b,\, k,\, \beta.\,$ Worthy of remark is the fact that the estimates (\ref{35}) hold for all $\,t,\,$ also when $\, T\, \rightarrow \infty.\,$


\begin {thebibliography}{99} 





\bibitem {i}Izhikevich E.M. {\it Dynamical Systems in Neuroscience: The Geometry of Excitability and Bursting}. The MIT press. England (2007)
\vspace{-3mm}
\bibitem {m2}J.D. Murray,   {\it  Mathematical Biology. II. Spatial models and biomedical applications }, Springer-Verlag, N.Y  2003 
\vspace{-3mm}
\bibitem {m1}J.D. Murray,   {\it  Mathematical Biology. I. An Introduction  }, Springer-Verlag, N.Y  2002 
\vspace{-3mm}
\bibitem{ks} J. P. Keener - J. Sneyd {\it  Mathematical Physiology }Springer-Verlag, N.Y  1998 
 \vspace{-3mm}

\bibitem{kss}Krupa, M; Sandstede, B; Szmolyan, P {\it Fast and slow waves in the FitzHugh-Nagumo equation.} J. Differential Equations 133 (1997), no. 1, 49--97.
\vspace{-3mm}


\bibitem {gj} L.Glass, M.E. Josephson {\it Resetting and annihilation of reentrant abnormally rapid heartbeat} Phys. Rev. Lett 75,10 (1995) 2059-2062
\vspace{-3mm}

 
\bibitem{r} J Rinzel {\it Models in neurobiology}  Nonlinear Phenomena in Physics and Biology. Edit by R. H. Enns, B. L. Jones, R. M. Miura, and S.nd S. Rangnekar. D. Reidel Publishing Company, Dordrecht-Holland. NATO Advanced Study Institutes Series. Volume B75, (1981), p.345 

 \vspace{-3mm}



\bibitem {nay}Nagumo, J., Animoto, S., Yoshizawa, S {\it An active pulse transmission line simulating nerve axon} (1962), Proc. Inst. Radio Engineers, 50, 2061-2070. 

\vspace{-3mm}


\bibitem {f61}FitzHugh R. {\it Impulses and physiological states in theoretical models of nerve membrane.} Biophysical journal 1. (1961), 445- 466.  
 \vspace{-3mm}




\bibitem{wzd}Wang, Jiang; Zhang, Ting; Deng, Bin  {\it Synchronization of FitzHugh-Nagumo neurons in external electrical stimulation via nonlinear control.} Chaos Solitons Fractals 31 (2007), no. 1, 30--38.  

\vspace{-3mm}
\bibitem {aa} J.G. Alford, G. Auchmuty {\it Rotating wave solutions of the FitzHugh-Nagumo equations.} J. Math. Biol.  53(2006) 797-819
\vspace{-3mm}
\bibitem{bcf06}Bini D., Cherubini C., Filippi S. {\it Heat transfer in FitzHugh-Nagumo models.} Physical Review E.  74, 041905 (2006)
\vspace{-3mm}
\bibitem {bcf} Bini, D.; Cherubini, C.; Filippi, S. {\it Viscoelastic  FitzHugh-Nagumo models.} Phys. Rev. E (3) 72 (2005), no. 4, 041905-01 12
\vspace{-3mm}

 
\bibitem  {rt}Rinzel, J; Terman, D {\it Propagation phenomena in a bistable reaction-diffusion system.} SIAM J. Appl. Math. 42 (1982), no. 5, 1111--1137.
 
\vspace{-3mm}
\bibitem{nya} J. Nagumo, S. Yoshizawa, Animoto, S. {\it Bistable Transmission Lines } IEEE Transactions on  Circuits and Systems, 12, Issue 3, (1965) Page(s): 400 - 412
\vspace{-3mm}







 






\bibitem {mk}   McKean, H. P., Jr. {\it  Nagumo's equation.} Advances in Math. 4 1970 209--223 (1970).
\vspace{-3mm}
\bibitem{rk}J Rinzel and J B Keller. {\it Traveling wave solutions of a nerve conduction equation.} Biophysical Journal, 13: 1313-1337, 1973. 
 \vspace{-3mm}

\bibitem {r}Rinzel J. {\it Spatial stability of traveling wave solutions of a nerve conduction equation. } Biophys J. 1975 15  975-988. 
\vspace{-3mm}
\bibitem{f}JA Feroe {\it Temporal stability of solitary impulse solutions of a nerve equation} Biophys. J. 1978 21 103-110.
\vspace{-3mm}
\bibitem {f82}J. A. Feroe,{\it Existence of travelling wave trains in nerve axon equations}, SIAM J. Appl. Math. 46 (1986), 1079-1097. 
\vspace{-3mm}
\bibitem {t}A. Tonnelier, {\it The McKean's caricature of the FitzHugh-Nagumo model. I : The space-clamped system }, SIAM Journal on Applied Mathematics 63, pp. 459-484 (2002)
\vspace{-3mm}


\end {thebibliography}
\end{document}